\documentclass[english]{lni}

\usepackage{booktabs}
\usepackage{url}
\usepackage{multirow}
\usepackage{subfig}
\usepackage{graphicx}
\usepackage{fancyhdr}
\usepackage{changepage} 
\usepackage{listings} 
\usepackage{color}
\usepackage{tablefootnote}

\usepackage[figurename=Fig., tablename=Tab.]{caption}

\fancypagestyle{titlepage}{
\fancyhead[RO]{\small F.  Boutros,  N. Damer,  M. Fang,  M. Gomez-Barrero,  K. Raja,  \linebreak C. Rathgeb,   A.  Sequeira,  and M. Todisco (Eds.): BIOSIG 2024, \linebreak Lecture Notes in Informatics (LNI), Gesellschaft f\"ur Informatik, Bonn 2024} 
\fancyfoot{}}

\renewcommand{\headrulewidth}{0.4pt} 
\author{Xuechen Liu\footnote{National Institute of Informatics,
2-1-2 Hitotsubashi, Tokyo 101-8430, Japan, \{xuecliu, wangxin, jyamagis\}@nii.ac.jp}, \ Xin Wang\footnotemark[1], \
Junichi Yamagishi\footnotemark[1]}

\title{A Preliminary Case Study on Long-Form In-the-Wild Audio Spoofing Detection}

\begin{document}

\maketitle

\renewcommand{\refname}{References}
\setcounter{footnote}{2} 
\thispagestyle{titlepage}
\pagestyle{fancy}
\fancyhead{} 


\fancyfoot{} 
\renewcommand{\headrulewidth}{0.4pt} 

\begin{abstract}
Audio spoofing detection has become increasingly important due to the rise in real-world cases. Current spoofing detectors, referred to as spoofing countermeasures (CM), are mainly trained and focused on audio waveforms with a single speaker and short duration. This study explores spoofing detection in more realistic scenarios, where the audio is long in duration and features multiple speakers and complex acoustic conditions. We test the widely-acquired AASIST under this challenging scenario, looking at the impact of multiple variations such as duration, speaker presence, and acoustic complexities on CM performance. Our work reveals key issues with current methods and suggests preliminary ways to improve them. We aim to make spoofing detection more applicable in more in-the-wild scenarios. This research is served as an important step towards developing detection systems that can handle the challenges of audio spoofing in real-world applications.
\end{abstract}

\section{Introduction}
The rapid advancement of audio \emph{deepfake} technology is the result of sophisticated deep learning techniques for highly realistic speech generation and has raised significant concerns both research-wise and social-wise \cite{cointelegraph2024, brewster2021}. This audio spoofing paradigm not only transcends language barriers but also enables the widespread dissemination of deceptive deepfake content, posing severe threats to information security. In response, the research community has made substantial progress in audio spoofing detection, notably through initiatives like the ASVspoof challenge series \cite{asvspoof2015, asvspoof2019, asvspoof2021_summary} and the Audio Deepfake Detection (ADD) challenge series \cite{add2022}. There are also companies who have been working on audio deepfake detection with commercial services, such as Nuance\footnote{https://www.nuance.com/omni-channel-customer-engagement/authentication-and-fraud-prevention/biometric-authentication.html} and PinDrop\footnote{https://www.pindrop.com/deepfake}. These efforts have initiated and accelerated the development of effective \emph{countermeasure} (CM) models, typically implemented as binary classifiers, to detect synthesized speech. 
These developments reflect the growing sophistication of audio anti-spoofing techniques and the corresponding need for increasingly advanced detection methods.

The datasets used in audio spoofing detection research have also evolved over time. For instance, the ASVspoof challenge series has been introducing new datasets with each edition. Notably, ASVspoof2019 \cite{asvspoof2019} has gained widespread adoption within the research community due to its comprehensive inclusion of various spoofing algorithms. Its successor, ASVspoof 2021 \cite{asvspoof2021_summary}, incorporated established audio compression techniques, encompassing both transmission-based (e.g., a-law, $\mu$-law, G.722) and media-based (e.g., mp3, opus) methods. Subsequent variants of this dataset and works developed in parallel have introduced partially-spoofed audio segments \cite{partialspoof2022, cai2023avdeepfake1mlargescalellmdrivenaudiovisual, alali2024rfpdatasetrealfake}, different compression artifacts such as neural-based codec models \cite{codecfake_dataset2024}, and different acoustic environments with noise and reverberation \cite{spoof_detection_noisy2016}. Multi-lingual setups extending beyond English \cite{mlaad} have been discussed, along with incorporating alternative media data such as video alongside audio \cite{fakeavceleb,cai2023avdeepfake1mlargescalellmdrivenaudiovisual}. 

However, the above works have primarily focused on audio data either generated or collected with constraints that make them isolated from real-world applications. With the constraint on acoustic conditions and artifacts having been addressed \cite{spoof_detection_noisy2016, asvspoof2021_summary}, two significant limitations persist: duration and multi-speaker audio. Existing datasets, often derived from acoustically clean datasets like VCTK \cite{vctk} and LJspeech\footnote{https://keithito.com/
LJ-Speech-Dataset}, typically feature short audio segments, ranging from less than 2 seconds to around 10 seconds. Consequently, most state-of-the-art CMs are trained on them \cite{aasist2022, lcnn}. Meanwhile, real-world fake speech (e.g., \cite{youtube2}), on the other hand, often has different characteristics. Such kind of speech audio tends to be longer than several seconds and performance of the state-of-the-art CMs in longer speech is unknown.\footnote{For example, the AV-Deepfake1M dataset \cite{cai2023avdeepfake1mlargescalellmdrivenaudiovisual} contains 10 to 30 seconds of deepfake video.} Dealing with such \emph{long-form} audio is not as simple as just segmenting it into shorter segments, because these clips may either be fully faked or contain a mix of real and fake segments, and sometimes involve multiple spoofing techniques, as well as overlapping between real and fake content, often featuring multiple speakers. These complex situations pose new challenges for deepfake detection. The disparity between research datasets and real-world conditions motivates the need for more representative studies in audio spoofing detection.

Therefore, our main focus is on long-form spoof audio. Such samples are either entirely spoofed or a combination of genuine and spoofed segments, characterized by extended duration and various audio processing, such as trimming, volume normalization, and audio compression with background noise. In this preliminary study, our methodology involves generating long-form variants by concatenating short-form samples, allowing for arbitrary proportions of genuine and spoofed content, containing more than one speakers, compression artifacts and acoustic conditions simultaneously. We hypothesize that the current CM models cannot generalize well to long-form audio with a mix of genuine and spoof audio, considering the fact that the current CMs are trained on short speech data with no intra-speech variability of the speaker or environment, we aim to provide insights into the limitations of existing approaches and potential strategies for improving their robustness in more complex, real-world-oriented audio spoofing detection.

\begin{figure}[t]
    \centering
    \includegraphics[width=\linewidth]{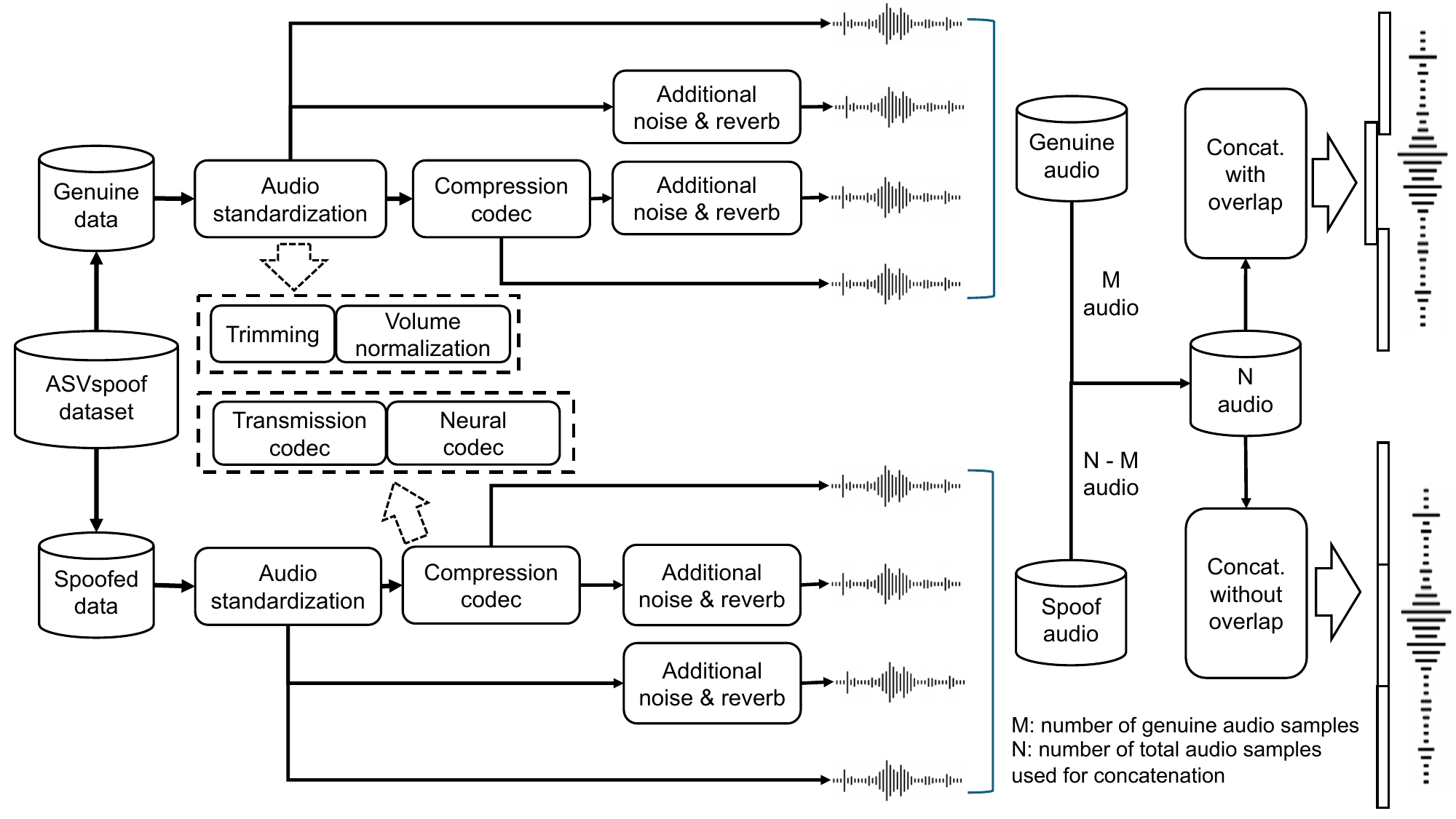}
    \caption{Audio generation and partitioning pipeline from short-duration audio to long-form audio. This can be applied to any audio dataset that contains genuine and spoof partitions. In this study, $N = 10$, and $M \in [0, 10]$. The compression codec and additional noise and reverberation effect can be found in Tab. \ref{tab:manipulation_methods}.}
    \label{fig:pipeline}
\end{figure}

\section{Long-form Spoofing Audio Generation}
As mentioned, our primary objective is developing long-duration spoofed audio with varying proportions of genuine and spoofed segments, containing multiple speakers, compression artefacts and acoustic conditions simultaneously.
Figure \ref{fig:pipeline} illustrates in detail the generation pipeline, which encompasses multiple stages of audio processing.

It is important to note that the binary ``genuine/spoof" labels assigned to the original short speech segments remain unchanged by any audio processing utilised prior to waveform concatenation in this study. This approach is intuitive and in line with our aim to detect speech outputs of generative models such as text-to-speech and voice conversion, no matter what kind of audio processing is performed.

\subsection{Processing Pipeline}
\label{secsec:pipeline}

\noindent 
\textbf{Source Audio Dataset.} The source audio dataset utilized in this study is the ASVspoof2019 \emph{logical access} (LA) track \cite{WANG2020101114}\footnote{https://datashare.ed.ac.uk/handle/10283/3336}. This open-sourced derivation from VCTK incorporates 17 distinct spoofing algorithms, with 19 spoofing attacks implemented. It features a well-structured data protocol comprising training, development, and evaluation sets. The training dataset encompasses 20 speakers and 6 synthesis algorithms selected from the total 19. The development set uses the same set of algorithms, while the evaluation set is constructed using the remaining 13 ones. This dataset has gained popularity in audio spoofing detection research, primarily due to the high quality and cleaner acoustic conditions offered by the original VCTK, as well as the sophisticated nature of the spoofing algorithms acquired. 

\noindent 
\textbf{Audio standardization}.
First, to address volume variations and silence in audio samples, we standardized the audio in two steps: trimming silence at the beginning and end of the audio using \emph{sox}\footnote{https://sourceforge.net/projects/sox}, followed by audio volume randomization to a certain range, within which the expected RMS value of the short audio waveform is randomly determined with ITU P.56 standard\footnote{https://github.com/openitu/STL}. The purpose of these is, respectively, to provide a fluctuation in the volume of the short audio without severe clipping and to avoid the long audio file to be generated being mainly occupied by silence.

\textbf{Audio compression and noise addition}.
Then, we use multiple audio processing techniques written in Table \ref{tab:manipulation_methods} to introduce various compression artefacts and acoustic environmental variations. Presented in Tab. \ref{tab:manipulation_methods} and Fig. \ref{fig:pipeline}, the audio processing methods applied can be categorized into three groups: traditional transmission-based and media compression ones, recently proposed neural-based compression ones, and additional noise and reverberation effect (via signal convolution).
Each standardized short audio sample is subjected to four variations for subsequent processing: (1) its original form, (2) a noise-added version, (3) a compressed version, and (4) a version with compression applied first, followed by noise addition\footnote{Preliminary experiments indicated that applying noise followed by compression is not a viable option.}. One of the compression methods was chosen at random, and noise and room impulse responses, also chosen at random from the respective databases \cite{musan, rir}, were used.

\begin{table}[t]
\centering
\caption{List of audio processing methods for this study. Transmission and media compression were implemented using \emph{sox}. All neural compression methods were applied with default parameter settings. The bitrate of each compression model is shown in the brackets in kilobits per second (kbps). The signal-to-noise ratio was 10dB when applying MUSAN.}
\label{tab:manipulation_methods}
\begin{tabular}{c|c}
\toprule
Audio processing & Employed technique/model \\
\midrule
Transmission \& media compression & mp3 (128), opus (64), a-law (64), $\mu$-law (64) \\ \hline
\multirow{2}{*}{Neural compression} & Soundstream (3) \cite{soundstream}, SpeechTokenizer (4.8) \cite{speechtokenizer}, \\ 
& EnCodec (24) \cite{encodec}, FACodec (4.8) \cite{naturalspeech3} \\ \hline
Additional noise and reverb & MUSAN \cite{musan}, Room impulse response \cite{rir} \\
\bottomrule
\end{tabular}
\vspace{-5mm}
\end{table}

\noindent 
\textbf{Long-form audio generation via concatenation.}
%
Illustrated at the right-hand side of Fig. \ref{fig:pipeline}, following earlier steps, we performed concatenation to generate long-form audio. We controlled the total number of audio segments used to generate a single long-form audio sample $N = 10$, and the number of genuine audio segments used in the generation $M \in [0, 10]$. For convenience of presentation and clarity, we define \emph{genuine ratio} as $M : (N - M)$.

The selection of audio segments for concatenation and their order within the long-form samples were determined randomly. Another variable in this process is the amount of overlap between adjacent audio segments. For this preliminary study, the overlap was implemented non-aggressively by overlaying the last 0.1 second of one segment with the first 0.1 second of the next, without cutting either, as shown on the rightmost side of Fig. \ref{fig:pipeline}.
Exploring more aggressive and randomized overlapping between adjacent audio segments remains a focus for future work.

Regarding the binary spoofing labels for the resulting long-form audio, if any spoof audio segments are involved in the concatenation (i.e., $0 < M < 10$), the entire long-form sample is labeled as \emph{spoof}. If no spoof segments are used (i.e., $M = 10$), the long-form sample is labeled as \emph{genuine}. This long-form audio database also allows for further research topics on spoofing such as spoofing localization, which is beyond the scope of this study.

\subsection{Differences from related databases}

One of the most relevant databases is the AV-Deepfake1M dataset \cite{cai2023avdeepfake1mlargescalellmdrivenaudiovisual}. The audio segments in this dataset are up to 30 seconds long, and include partially synthesized speech based on voice cloning. The synthetic speech segments in the long-form speech are quite short, with an average ratio of about 4\%. Moreover, it includes only two synthesis methods. In contrast, our dataset features a variable fake speech ratio, with a wider variety of synthesis methods. Additionally, our dataset differs from the RFP dataset \cite{alali2024rfpdatasetrealfake}, which combines synthetic speech with human speech and various noises but does not account for reverberation, compression, or overlap. \cite{newapproachvoiceauthenticity} introduces multiple voice editing techniques to create audio, but it does not include long-form audio, and the objective of it is to identify the type of audio processing, while part of the aim of our study is towards a robust audio deepfake detector to various types of audio processing.

\begin{figure}[!h]
\centering
\subfloat[\texttt{Baseline pre-trained model \cite{aasist2022}}\label{fig7:baseline}]{
\includegraphics[width=0.48\linewidth]{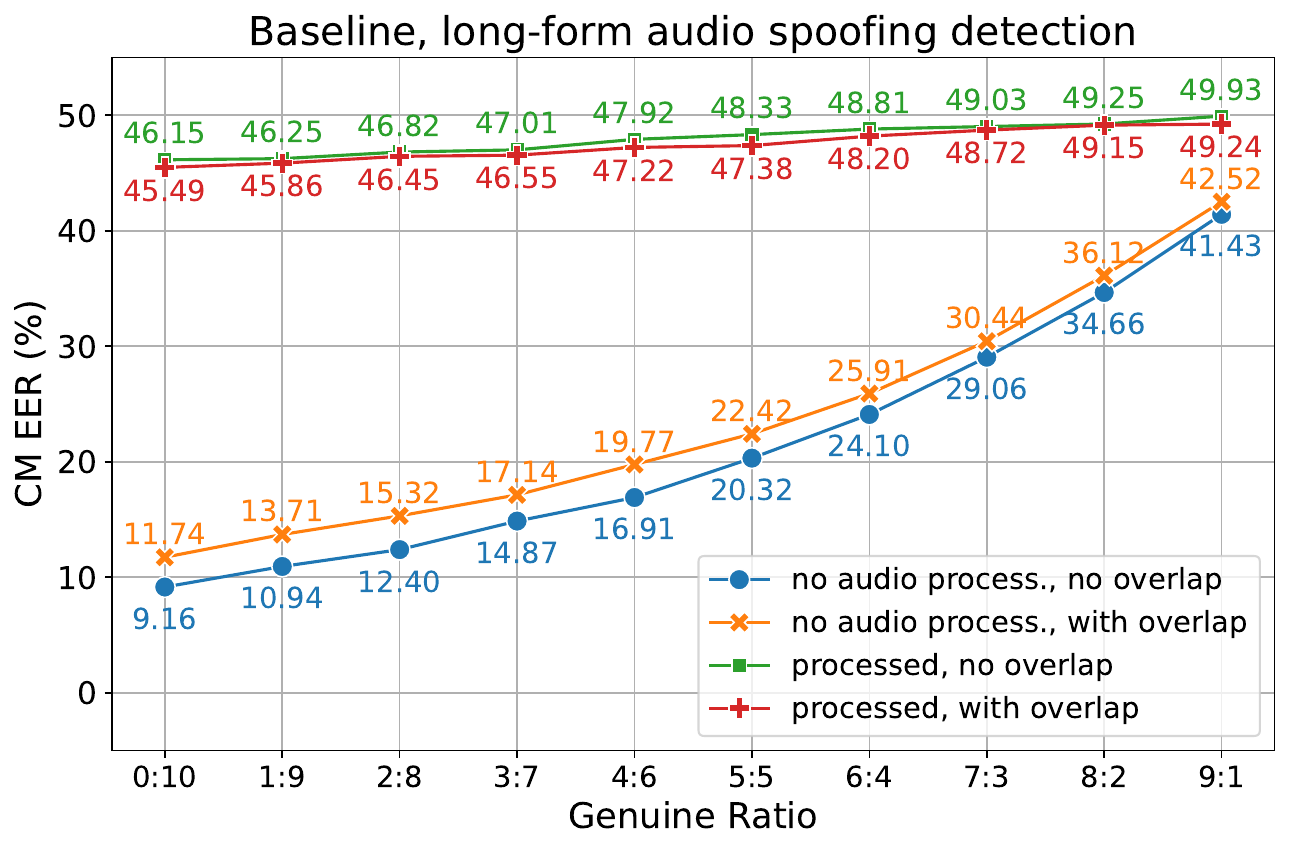} 
}
\subfloat[\texttt{B01}\label{fig7:b01}]{
\includegraphics[width=0.48\linewidth]{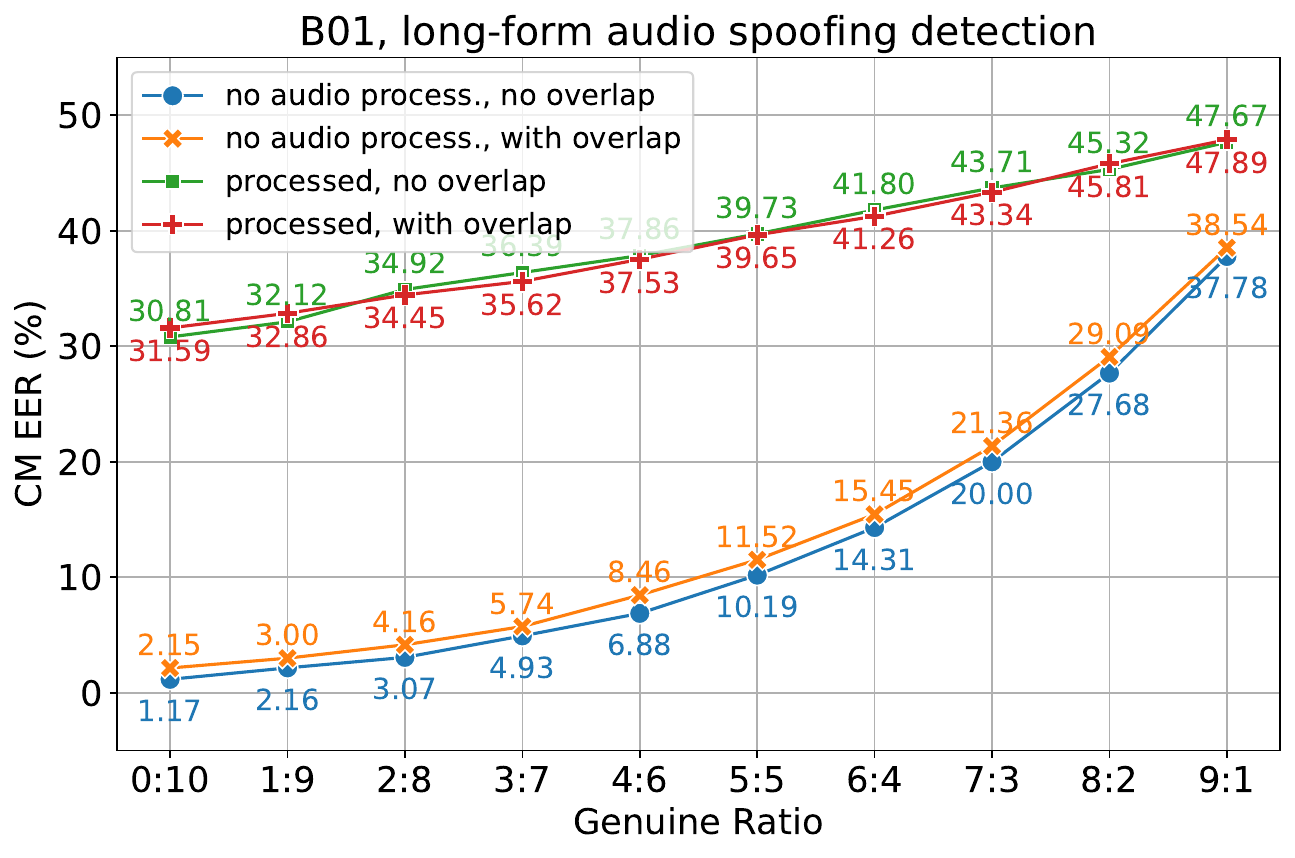} 
} \par \vspace{-3mm}
\subfloat[\texttt{B02}\label{fig7:b02}]{
\includegraphics[width=0.48\linewidth]{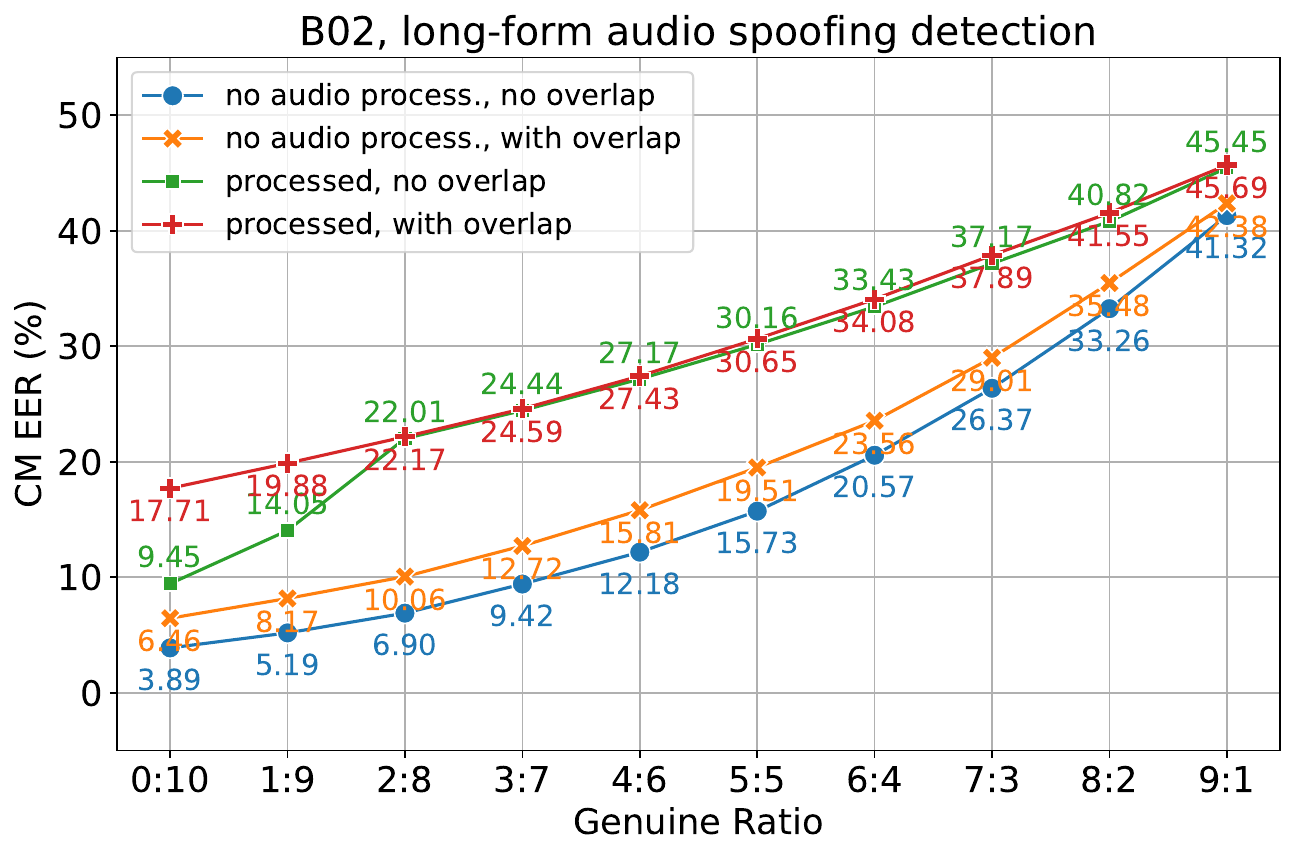} 
} 
\subfloat[\texttt{B03}\label{fig7:b03}]{
\includegraphics[width=0.48\linewidth]{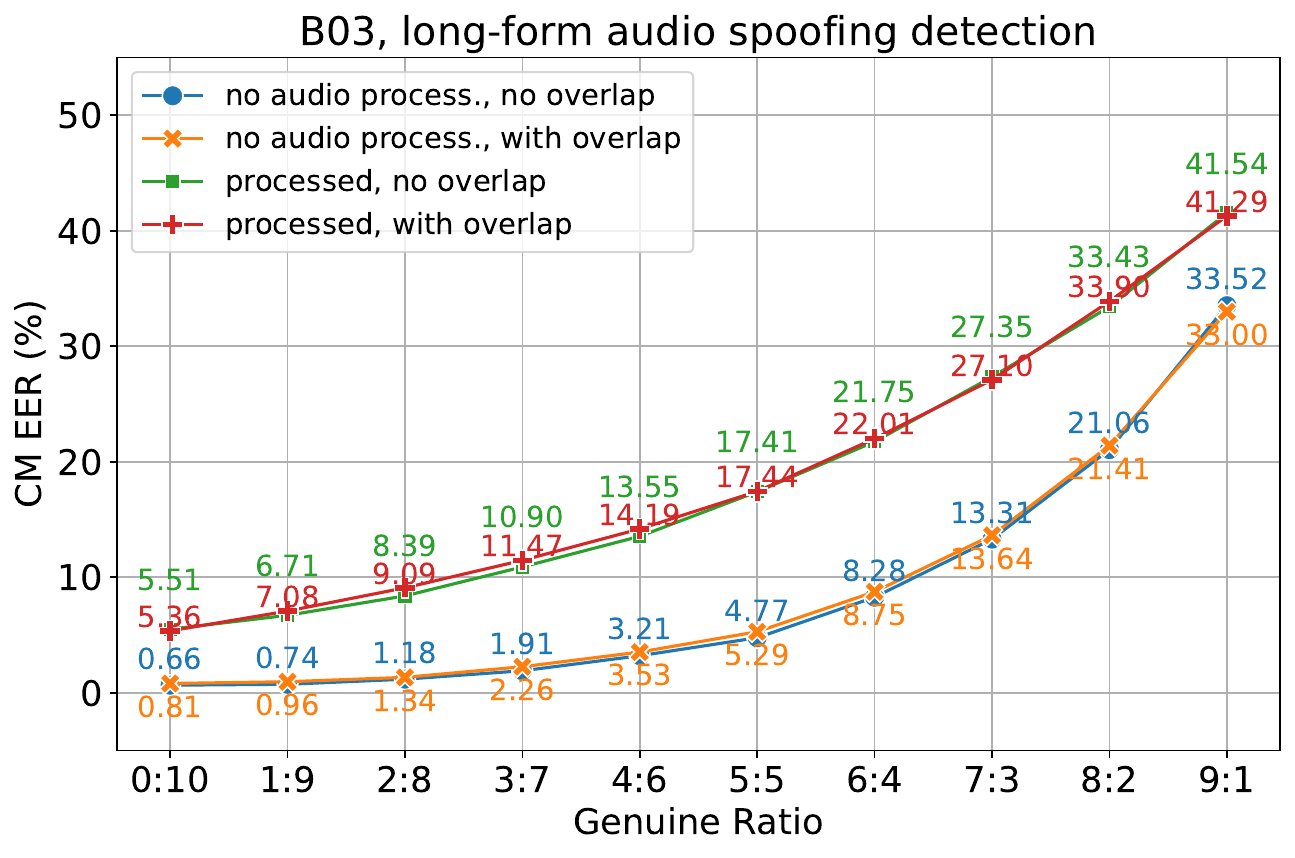} 
} \par \vspace{-3mm}
\subfloat[\texttt{B04}\label{fig7:b04}]{
\includegraphics[width=0.48\linewidth]{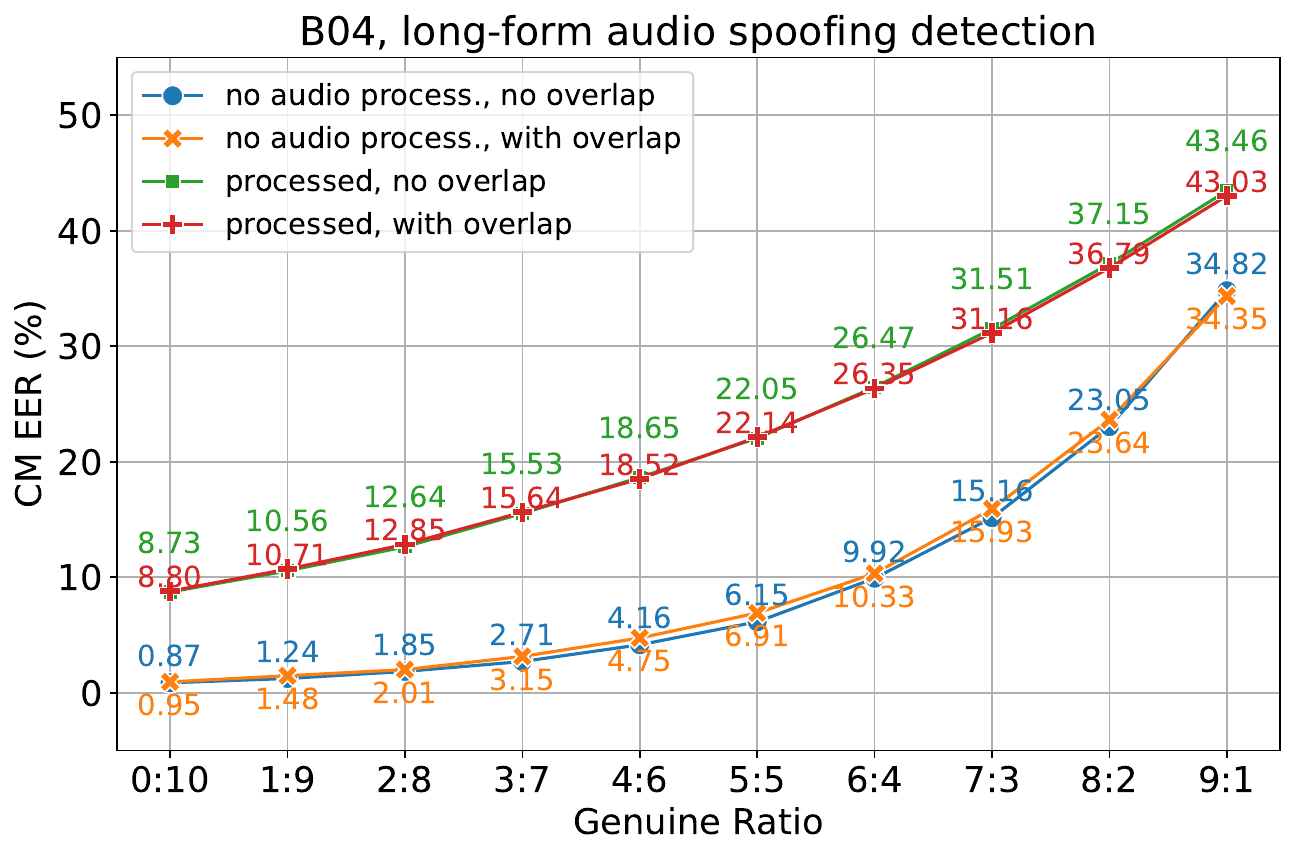} 
}
\subfloat[\texttt{B03, with score-level averaging}\label{fig7:b03_segmented}]{
\includegraphics[width=0.48\linewidth]{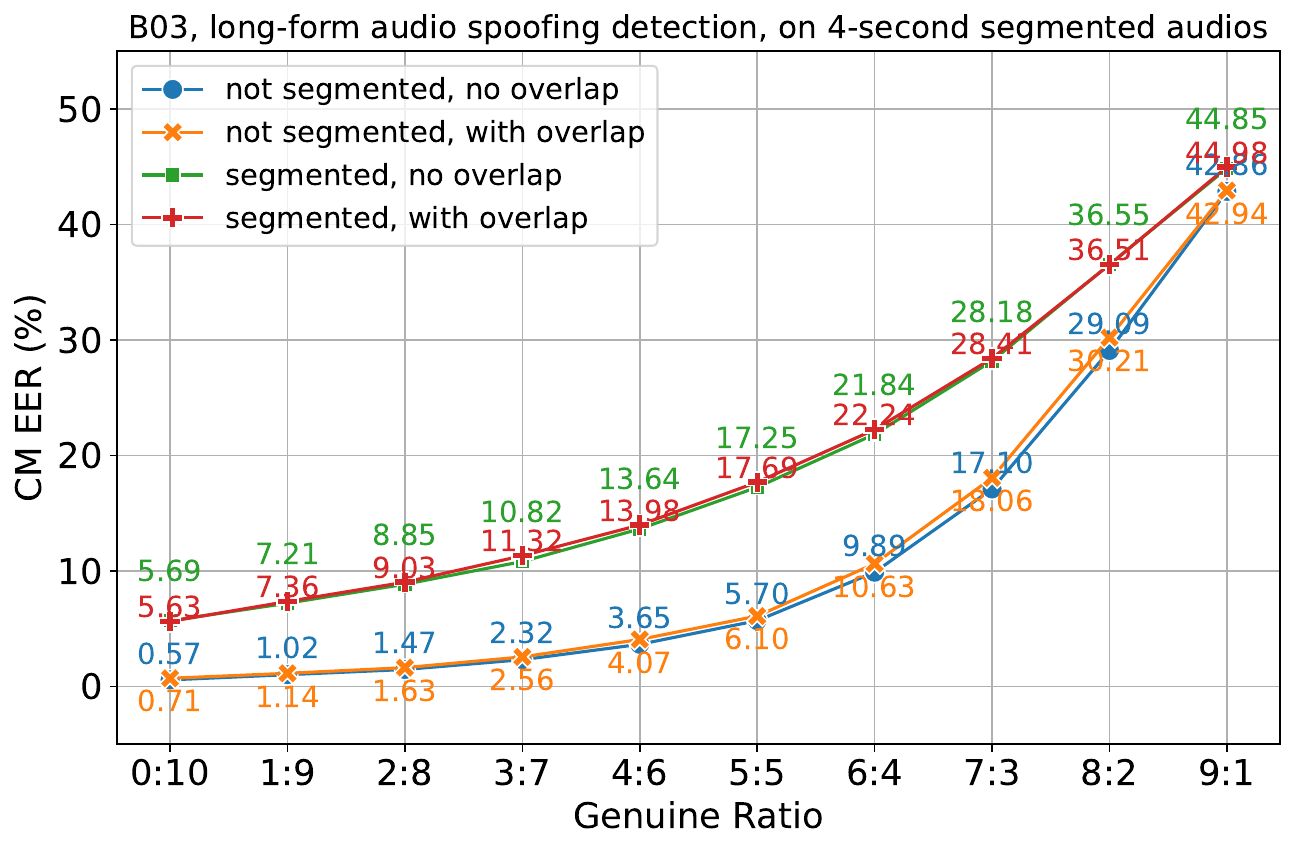} 
}
\caption{Performance of trained systems on generated long-form audio with various audio processing and overlap between adjacent source audio segments. The ``audio process" in the figure represents the neural codec compression and adding noise and reverberation. Note that the x-axis of each figure shows the genuine ratio of \emph{spoofed} audio. Genuine ratio of 10:0 corresponds to \emph{genuine} audio.}
\label{fig:performances}
\end{figure}

\section{Experimental Setup}
\label{sec:exp_setup}

\textbf{Training data and experimental CMs}.
Using the source ASVspoof2019 LA training set and the proposed pipeline described in Section \ref{secsec:pipeline}, we created and compared multiple training sets to testify the effectiveness of the proposed pipeline. They are listed below.
\begin{itemize}
\setlength{\parskip}{-2mm}
    \item Training set 1 was created on the basis of the source data set with audio standardization, without additional processing (compression, noise, reverb) or producing long-form audios;
    \item Training set 2 was created with audio standardization and processing but not long-form audios;
    \item Training set 3 was created with audio standardization and processing. All the data are concatenated long-form audios without overlapping;
    \item Training set 4 was similar to training set 3 except that long-form audios generation was done with 0.1s overlapping.
\end{itemize}
The \emph{source} ASVspoof2019 LA training set was also used in the experiments for reference. All training sets, except Train set 1 and source set, have 2,580 genuine and 22,800 spoofed long-form audio files each, and they are subsampled from the corresponding long-form audio sets; Train set 1 and source set contain the same number of audio files but are in short-form.
These training sets led to five CMs: \texttt{B01}, \texttt{B02}, \texttt{B03}, and \texttt{B04} trained on the training sets 1, 2, 3, and 4, respectively, and \texttt{Baseline} trained on the source data set.
All of them used the AASIST \cite{aasist2022}  architecture and were implemented based on an open-source repository\footnote{https://github.com/clovaai/aasist}. The training procedure hyper-parameters adhered to those specified in \cite{aasist2022}. Note that following the standard setup described, at training stage the input waveform was always chunked to 4 seconds, while at evaluation stage the CM model accepted input with varied length.
The prediction of the AASIST model for long speech can thus be done in two ways. The first is to process all long speech as input and output a single score, while the second is to segment it into chunks of the same length as the training (4 seconds) and merge multiple scores from each chunk. If not explicitly stated, the first method is used in this paper.

\textbf{Evaluation data}. 
Performance of each trained CM was measured on four long-form audio evaluated sets created using the proposed pipeline and the source ASVspoof2019 evaluation set, regardless of whether the CM training data was short or long-form speech. Specifically, with the standardization having processed, the four evaluation sets vary in terms of two configurations in the processing pipeline: 1) whether the source audio segments used for concatenation (both genuine and spoof) were processed, marked as ``no audio process." or ``processed", and 2) whether there was overlap when concatenating the short audios, marked as ``no overlap" or ``with overlap".
Each evaluation set contains 10,000 genuine and 20,000 spoofed long-form audio files, respectively, and contains variants for each genuine ratio, ranging from $0:10$ to $9:1$.

\section{Results \& Discussion}
\label{sec:results}
The results are shown in Fig. \ref{fig:performances}, from which we have a few findings.

\textbf{CM trained on short-form data cannot generalize well to long-form test data}. This is supported by the result that \texttt{Baseline}, which obtained 0.83\% equal error rate (EER) on the ASVspoof2019 evaluation set, performed much worse on the four evaluation sets created in this study. As Fig. \ref{fig:performances} (a) shows, on the long-form audio evaluation set without audio processing, either with overlap or not, the EERs became higher than 9\%. If the evaluation set went through audio processing, the EERs became higher than 45\%. This finding adheres to earlier works on recent audio deepfake datasets.\footnote{\cite{alali2024rfpdatasetrealfake} and \cite{codecfake_dataset2024} returned 29.01\% and 34.99\% EER with AASIST, respectively.}

\textbf{More generalizable CMs need long-form training data}. 
Compared with \texttt{Baseline}, the performance of \texttt{B01} and \texttt{B02} improved to different degrees on the evaluation sets across the varied genuine ratios. A potential reason is that the audio standardization and various audio processing applied to the training data helped to mitigate the training-evaluation mismatch. However, more improvements were found in the cases of \texttt{B03} and \texttt{B04}, where both used long-form audios in the training set. In particular, when the genuine ratio is 0:10, the EERs of \texttt{B03} were lower than 6\%. Given that modern neural-based CM classifiers are predominantly trained on short, single-speaker audio segments, this finding highlights the need to incorporate longer-duration audio samples with diverse speakers and acoustic conditions into the training data.

\textbf{Effect of genuine ratio}. As illustrated in all sub-plots of Fig.~\ref{fig:performances}, across all the four evaluation sets, the EER increases as the genuine ratio changes from $0:10$ to $9:1$, where the proportion of source genuine short segments increased. This trend holds even for CMs \texttt{B03} and \texttt{B04}, which were trained using long-form audios. This finding is analogously consistent with \cite{partialspoof2022} where very short synthetic segment was inserted into a short real speech although our audio is longer and contains longer synthetic segment.

\textbf{Effect of overlap}. It is interesting that applying pre-settled 0.1-second overlap between adjacent audio segments during concatenation to the evaluation set does not always make data more challenging. When no audio processing was done on the evaluation data, the EERs of \texttt{Baseline}, \texttt{B01} and \texttt{B02} on the evaluation set with the overlap are higher than their counterparts without overlap (i.e., comparing the yellow and blue dots in Fig.~\ref{fig:performances}(a), (b), and (c)). 
However, for \texttt{B03}, the effect of the overlap in the evaluation set is less notable. This discrepancy may have arisen because the overlapped portions of long-form audio sometimes contain multiple speakers, potentially mixed with the aforementioned variations (particularly noise), leading to confusion in CMs trained on such data. These observations necessitates further research on localization techniques that account for spoofing information in such scenarios. Meanwhile, adding the same overlap ingredient into the training data, resulting in \texttt{B04}, did not lead to better CM performance compared with \texttt{B03}. This is worth further investigation to determine whether overlapping is beneficial and can generate proper overlapping spoof audio for effective CM training.

\textbf{A different way of scoring the evaluation data}. 
As described before, the prediction of the AASIST model for long speech can be done using multiple segmented audio.
To investigate the impact on different ways of scoring, we segmented the long-form utterances into 4-second chunks and evaluated them using \texttt{B03}, the best-performing model. We then averaged the scores to return a single CM score for each long-form utterance. The results are shown in Fig. \ref{fig7:b03_segmented}. Comparison with \ref{fig7:b03} shows that segmenting long-form audio, followed by averaging at inference time, improves performance without audio processing and with a low genuine ratio, but slightly worsens performance in other cases. This suggests that simply segmenting long audio does not notably change performance, indicating a need for architectures and scoring methods suitable for long-form audio.

\section{Conclusion}
This paper presents a preliminary case study on audio deepfake detection, addressing limitations in long-duration spoofed audio with varying proportions of genuine and spoofed segments, containing multiple speakers, compression artefacts and acoustic conditions simultaneously. We have extended the ASVspoof2019 LA dataset, generating variants with various audio processing methods applied, longer audio duration and multiple speakers included. We have also included their variants with overlapped segments between adjacent source audio segments. 
We have incorporated the generated long-form audio into both evaluation and training stages, analyzing its impact on the CM performance. We examined the effect of extended duration, audio overlapping, and the ratio of genuine and spoofed content on detection performance. While the long-form evaluation data have been challenging for CMs trained on original short-duration audio segments, incorporating spoofed audios with longer duration have improved the CM performance substantially. Future research may focus on generating long-form spoofed audio using a wider range of compression and audio processing algorithms, along with developing robust spoof localization methods.

\section{Acknowledgements}
This study is partially supported by JST CREST Grants (JPMJCR18A6, JPMJCR20D3), JST AIP Acceleration Research (JPMJCR24U3), and MEXT KAKENHI Grants (24H00732). This research was also partially supported by the project for the development and demonstration of countermeasures against disinformation and misinformation on the Internet with the Ministry of Internal Affairs and Communications of Japan. This study was carried out using the TSUBAME4.0 supercomputer at Tokyo Institute of Technology.

\bibliography{lniguide}

\end{document}